\documentclass{PoS}

\usepackage{amsmath}
\usepackage{axodraw}
\usepackage{epsfig}

\newcommand{\be}{\begin{equation}}
  \newcommand{\ee}{\end{equation}}
\newcommand{\bea}{\begin{eqnarray}}
  \newcommand{\eea}{\end{eqnarray}}

\newcommand{\order}[1]{{\mathcal O}\left(#1\right)}
\newcommand{\scs}{\scriptscriptstyle}

\newcommand{\Leff}{{\mathcal L}_{\rm eff}}
\newcommand{\LQCDQED}{{\mathcal L}_{{\rm QCD} \times {\rm QED}}
  (u, d, s,c, b)}

\title{NNLO QCD corrections to the $\bf{ m_c}$ dependent 
       matrix elements in {\boldmath $ \overline B \to X_s \gamma$}}

\ShortTitle{$\order{\alpha_s^2}$ corrections to 
            $\langle s \gamma|Q_{1,2}|b \rangle $  
              in $ \overline{B} \to X_s \gamma$}

\author{\speaker{Radja Boughezal}\\
        Institut f\"ur Theoretische Physik
        und Astrophysik, Universit\"at W\"urzburg, \\
        Am Hubland, D-97074 W\"urzburg, Germany\\
        E-mail: \email{radja.boughezal@physik.uni-wuerzburg.de}}

\abstract{Recent developements in the calculation of the NNLO QCD corrections
          to the charm quark mass dependent matrix elements 
          in $ \overline{ B} \to X_s \gamma$ are reported.
          Special emphasis is put on the new results 
          of the  virtual $\order{\alpha_s^2}$ fermionic contribution
          to these matrix elements~\cite{Boughezal:2007ny}.}

\FullConference{8th International Symposium on Radiative Corrections (RADCOR)\\
		 October 1-5 2007\\
		 Florence, Italy}

\begin{document}
\section{Introduction}
The inclusive rare $ \overline{ B} \to X_s \gamma$ decay is a natural framework
for high precision studies of FCNC, thanks to its low sensitivity 
to non-perturbative effects. As a loop induced process 
in the Standard Model (SM), it is highly sensitive 
to new physics~\cite{Haisch:2007ic}.
In order to obtain stringent constraints on extensions of the SM
from this decay, accurate measurments and precise theoretical predictions
with a good control of perturbative and non-perturbative corrections 
have to be provided. \\
On the experimental side, the latest measurements by CLEO, Belle
and BaBar~\cite{experiments} have been combined by the Heavy Flavor 
Averaging Group into the current world average (WA) that reads 
for a photon energy cut of $E_{\gamma}> 1.6$ GeV~\cite{Barberio:2007cr} 
in the $ \overline{B}$-meson rest-frame
\be \label{eq:HFAG}
{\mathcal B}(\overline {B} \to X_s \gamma)_{\scs E_{\gamma} > 1.6\,{\rm GeV}}^{\scs\rm
exp}
= \left(3.55\pm 0.24{\;}^{+0.09}_{-0.10}\pm0.03\right)\times 10^{-4},
\ee
where the first error is given by the statistic and systematic uncertainty,
the second one is due to the theory input on the shape function, and the third
one is caused by the $b \to d\gamma$ contamination\,. This average
is in good agreement with the recent theoretical
estimate including known next-to-next-to-leading-order (NNLO) effects
~\cite{Misiak:2006zs}
\be
\label{theoretical B}
{\mathcal B}({\overline {B}}\to X_s\gamma)_{\scs E_{\gamma} > 1.6\,{\rm GeV}}^{\scs\rm theo} = (3.15 \pm 0.23) \times 10^{-4},
\ee
where the error consists of four types
of uncertainties added in quadrature: non-perturbative (5\%),
parametric (3\%), higher-order (3\%) and $m_c$-interpolation ambiguity (3\%).
The total error of the present experimental WA of about 7\% in 
Eq.~(\ref{eq:HFAG}) is expected to be reduced at future B factories
to approximately 5\%. In view of this accuracy, SM calculations need 
to be improved with the same precision level by completing 
the NNLO QCD program. \\
QCD corrections to the partonic decay rate
$\Gamma\left( b \to s \gamma\right) $ contain large logarithms
of the form \\ $\alpha_s^n\left(m_b\right) \, \ln^m\left(m_b/M_W\right)$,
with $m\le n$, which should be resummed 
with the help of renormalization-group techniques. A convenient framework
is an effective low-energy theory obtained from the SM by decoupling
the heavy electroweak bosons and the top quark\,. The resulting effective
Lagrangian is a product of
the Wilson coefficients $C_i(\mu)$ with local flavor-changing 
operators $Q_i(\mu)$\, up to dimension six.\\
A consistent calculation of $ b \to s \gamma$ at the NNLO level
requires three steps: $i)$ evaluation of $C_i (\mu_0)$ at the matching 
scale $\mu_0 \sim M_W$ by requiring
equality of Green's functions in the full and the effective theory up
to leading order in (external momenta)/$M_W$ to $\order{\alpha_s^2}$.
All the relevant Wilson coefficients have already been
calculated~\cite{Bobeth:1999mk,Misiak:2004ew} to this precision,
by matching the four-quark operators $Q_1,\dots,Q_6$ and the dipole operators
$Q_7$ and $Q_8$ at the $2$- and $3$-loop level respectively.
$ii)$ calculation of the operator mixing under renormalization, by
deriving the effective theory Renormalization Group Equations (RGE) and
evolving $C_i(\mu)$ from $\mu_0$ down to the low-energy
scale $\mu_b \sim m_b$, using the anomalous-dimension matrix (ADM)
to $\order{\alpha_s^3}$\,.
Here, the $3$-loop renormalization in the $\{Q_1, \dots ,Q_6\}$ and
$\{Q_7, Q_8 \}$ sectors was found in~\cite{Gorbahn:2004my,Gorbahn:2005sa},
and results for the $4$-loop mixing of $Q_1, \dots, Q_6$ into
$Q_7$ and $Q_8$ were recently provided in~\cite{Czakon:2006ss}, thus
completing the anomalous-dimension matrix.
$iii)$ determination of the on-shell matrix elements of the various
operators at $\mu_b~\sim~m_b$ to $\order{\alpha_s^2}$. 
This task is not complete yet, although a number of contributions is known.
The $2$-loop matrix element of the photonic dipole operator $Q_7$, 
together with the corresponding bremsstrahlung, was found 
in~\cite{Melnikov:2005bx,Blokland:2005uk}, confirmed in~\cite{Asatrian:2006ph} 
and subsequently extended to include the full charm quark mass 
dependence in~\cite{Asatrian:2006rq}. In~\cite{Bieri:2003ue}, 
the $\order{\alpha_s^2\,n_f}$ contributions were found to the $2$-loop 
matrix elements of $Q_7$ and $Q_8$, as well as to the $3$-loop matrix 
elements of $Q_1$ and $Q_2$, using an expansion in the quark mass 
ratio $m_c^2/m_b^2$.
Diagrammatically, these parts are generated by inserting a $1$-loop quark
bubble into the gluon propagator of the $2$-loop Feynman diagrams.
Naive non-abelianization (NNA) is then used to get an estimate
of the complete corrections of $\order{\alpha_s^2}$ by replacing
$n_f$ with $-\frac{3}{2}\beta_0$. Moreover, the contributions of the
dominant operators at $\order{\alpha_s^2 \beta_0}$ to the photon energy 
spectrum have been computed in~\cite{Ligeti:1999ea}.

A rather important and difficult piece that is still missing 
to date is the complete $\order{\alpha_s^2}$ calculation of 
the matrix elements of the four-quark operators $Q_1$ and $Q_2$.  
These operators contain the charm quark, and the main source 
of uncertainty at the NLO level is related to the ambiguity 
associated to the choice of scale and scheme for $m_c$~\cite{Gambino:2001ew}.
As these matrix elements start contributing for the first time 
at $\order{\alpha_s}$, the choice of scale and scheme for $m_c$
is a NNLO effect in the branching ratio. 
Therefore a calculation of $\langle s \gamma|Q_{1,2}|b \rangle $ 
at $\order{\alpha_s^2}$ is crucial to reduce the overall theoretical 
uncertainty in ${\mathcal B}({\overline B}\to X_s\gamma)$ . 
In~\cite{Misiak:2006ab}, the full matrix elements of $Q_1$ and
$Q_2$ have been computed in the large $m_c$ limit, $m_c\gg
m_b/2$. Subsequently, an interpolation
in the charm quark mass has been done down to the physical region,
under the assumption that the $\beta_0$-part is a good approximation at
$m_c\,=\,0$\,. This is the source of the interpolation uncertainty
mentioned below Eq.~(\ref{theoretical B}).
Reducing this uncertainty requires the evaluation of the $3$-loop
$\langle s \gamma|Q_{1,2}|b \rangle $ at $m_c=0$, whereas removing it
involves their calculation at the physical value of $m_c$, namely dealing
with hundreds of $3$-loop on-shell vertex diagrams with two scales $m_b$
and $m_c$ which is a formidable task. Both of these calculations are
being pursued in~\cite{O2BSG}, and we will comment on the current status 
in the next section. An important subset of diagrams contributing 
to the virtual $3$-loop on-shell calculation of 
$\langle s \gamma|Q_{1,2}|b \rangle $ for 
$m_c \neq 0$ is the fermionic part which
constitutes a major input both for the NNA and for the interpolation 
of the non-NNA terms between $m_c \gg m_b/2$ and $m_c < m_b/2$, 
and are thus crucial for the accuracy of Eq.~(\ref{theoretical B}).
A result for these diagrams was presented in~\cite{Bieri:2003ue}
assuming that $n_f \,=\, 5$ massless fermions are present in the quark loop 
inserted into the gluon propagator. An independent check of this calculation
as well as the validity of the massless approximation,
and new results for the missing diagrams with heavy $b$ and $c$ 
quark loops have been recently given~in~\cite{Boughezal:2007ny}.  
\section{Calculation of the matrix elements 
$\langle s \gamma|Q_{1,2}|b \rangle $}
\begin{figure}
  \begin{center}
    \epsfig{file=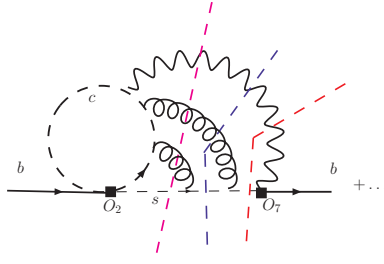, width=.35\textwidth}
    \caption{\label{sample CutQ2}\sf An example graph of a cut $4$-loop 
      b-quark self-energy. \sf}
  \end{center}
\end{figure}

The $\order{\alpha_s^2}$ calculation of the matrix elements
$\langle s \gamma|Q_{1,2}|b \rangle $ is done within the framework
of an effective theory with the Lagrangian 
\be \label{eq:effectivelagrangian}
\Leff = \LQCDQED + \frac{4G_F}{\sqrt{2}} V^\ast_{ts} V_{tb}
\sum^{8}_{i= 1} C_i (\mu) \, Q_i (\mu).
\ee
Adopting the operator definitions of~\cite{Chetyrkin:1996vx},
the physical operators that are relevant for our calculation together
with the size of their Wilson coefficients read
\be \label{eq:operators}
\begin{array}{l@{\hspace{5mm}}l} Q_{1,2} = 
  (\overline{s} \Gamma_i c)(\overline{c} \Gamma^\prime_i b) \, , & 
       C_{1,2} (m_b) \sim 1 \, , \\[1.5mm]
  Q_{3 \text{--} 6} = (\overline{s} \Gamma_i b) \sum_q (\overline{q}
        \Gamma^\prime_i q) \, , &
        \left | C_{3 \text{--} 6} (m_b) \right | < 0.07 \, , \\[1.5mm]
  Q_7 = e/g_s^2\;{\overline m_b}(\mu)\, (\overline{s}_L
        \sigma^{\mu \nu} b_R) \, F_{\mu \nu}  \, , &
        C_7 (m_b) \sim -0.3 \, , \\[1.5mm]
  Q_8 = 1/g_s\;{\overline m_b}(\mu)\, (\overline{s}_L
        \sigma^{\mu \nu} T^a b_R) \, G^a_{\mu \nu}  \, , & 
        C_8 (m_b) \sim -0.15 \, ,
\end{array}
\ee
where $\Gamma$ and $\Gamma^\prime$ stand for various products 
of Dirac and color matrices. A possible way of getting the complete
matrix elements at $m_c=0$ is by interfering the operators $Q_1$ and $Q_2$
with the magnetic dipole operator $Q_7$, then cutting the resulting
$4$-loop propagator diagrams in all possible ways that contain 
a photon and an s-quark in the final state. 
In total, $506$ diagrams are generated this way
each of which involves up to $5$-particle cuts if final states with
$c\overline{c}$ production are considered. A sample graph
is shown in FIG.~\ref{sample CutQ2}.   
Since charmed hadrons in the final state are excluded experimentally,
the ${\mathcal B}({\overline{B}}~\to~ X_s\gamma)$ does not contain 
contributions from $c\overline{c}$ production. Thus, a perturbative 
calculation of $ b~\to~ X_s^{\scs{parton}}\gamma $ should be done accordingly.
In order to avoid logarithmic divergences resulting at $m_c=0$ from $\ln\, m_c$
terms, cuts through the $c$-quark loop inserted into the gluon propagator
have been kept in our calculation. Their contribution will be subtracted
at the measured value of $m_c$ after performing the interpolation.
Moreover, since only the real part of the interference between
the matrix elements of $Q_2$ and $Q_7$ contributes to the decay of
$ b~\to~ X_s^{\scs{parton}}\gamma $, we do not distinguish between
masters that differ only in their imaginary part. This reduces
the number of masters to less than $200$.
Details related to their calculation will be given 
elsewhere~\cite{O2BSG}.\\
As far as the $\order{\alpha_s^2}$ $3$-loop virtual correction 
to $\langle s \gamma|Q_{1,2}|b \rangle $ at $m_c\neq 0$ is concerned,
the generated $420$ vertex diagrams have been expressed through 
$21231$ scalar integrals that depend on the scales $m_b$ and $m_c$.
With the help of Laporta algorithm~\cite{Laporta:2001dd}, 
they have been subsequently reduced to $476$ masters. 
The latter are being evaluated using a combined approach, namely Mellin-Barnes
technique together with differential equations solved numerically.
The same techniques have been applied in the calculation of the fermionic 
contribution, therefore we refer the reader 
to~\cite{Boughezal:2007ny,Boughezal:2006uu} for all related details. 
\section{Results for the fermionic diagrams}
\begin{figure}
  \begin{minipage}{1.\textwidth}
    \begin{center}
      \begin{minipage}{.49\textwidth}
        \epsfig{file=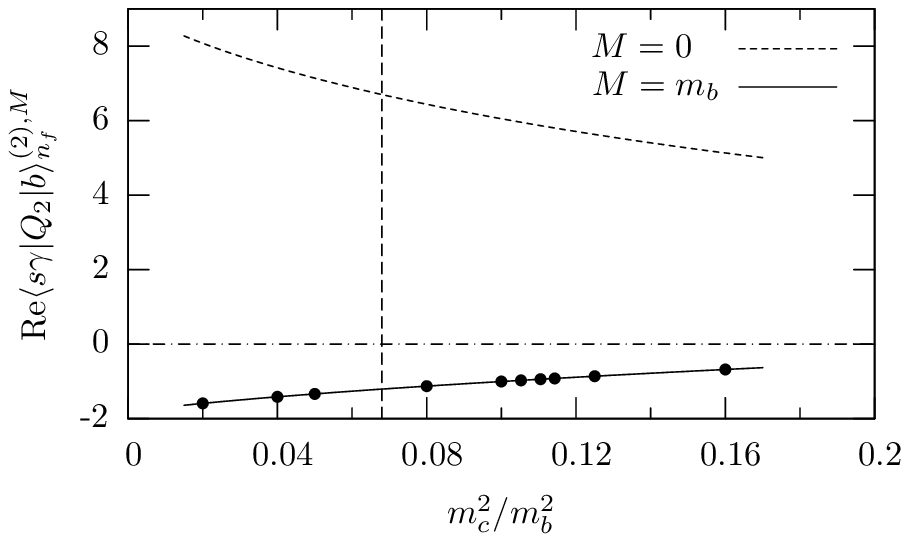, width=1.\textwidth}
        \centerline{(a)}
      \end{minipage}
      \begin{minipage}{.49\textwidth}
        \epsfig{file=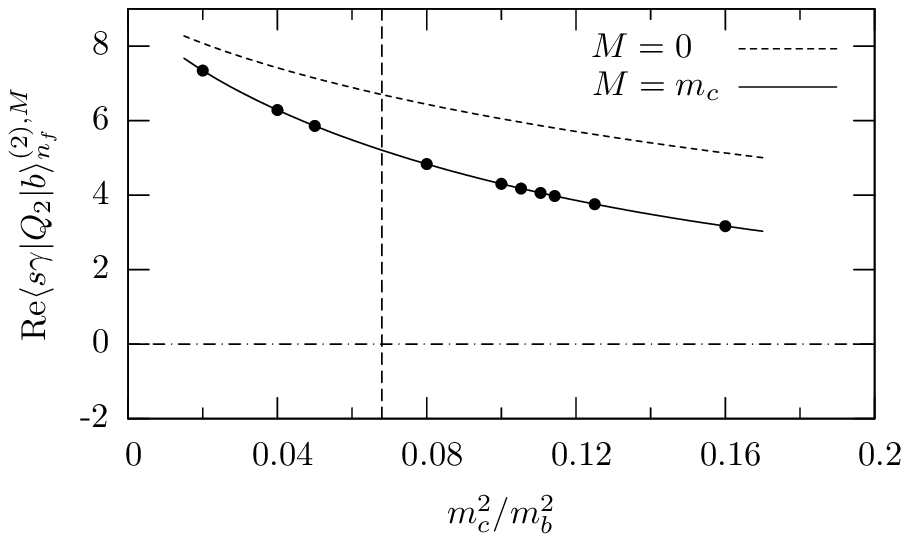, width=1.\textwidth}\hfill
        \centerline{(b)}
      \end{minipage}
    \end{center}
  \end{minipage}
  \caption{\label{Zdependence}\sf Plots of $\mbox{Re}\langle
    s\gamma|Q_2|b\rangle^{(2),M}_{n_f}$ as
    function of  $m_c^2/m_b^2$ with $M=m_b$ (a) and
    $M=m_c$ (b)~and~($\mu_b=m_b$, $ n_f=1$). 
    For comparison, we also show the $M=0$ case.\sf
}
\end{figure}

As was mentioned in the introduction, we have calculated 
the fermionic diagrams with three different quark loop insertions 
into the gluon propagator, namely a massless 
as well as heavy $b$ and $c$ quark loops. Since the massless
case was discussed in detail in~\cite{Bieri:2003ue}, 
we constrain ourself here to the new results related to the missing 
contributions from heavy loops and compare them with 
the massless approximation results. 
As, at ${\mathcal O}(\alpha_s^2 n_f)$, the matrix elements of $Q_1$ and $Q_2$
are related to each other by $\langle s \gamma|Q_{1}|b \rangle  = 
-{1}/{(2\, N_c)} \langle s \gamma|Q_{2}|b \rangle$,
we just give results for the matrix elements of $Q_2$. 
The normalization of our amplitude is defined as follows
\be
\langle s\gamma|Q_2|b\rangle_{{\mathcal O}(\alpha_s^2 n_f)} =
\left(\frac{\alpha_s}{4\pi}\right)^2\,\frac{e}{8\pi^2}\,m_b\,\,n_f\,
\langle s\gamma|Q_2|b\rangle^{(2),M}_{n_f}\,\,
\overline{u}_s \, R\,\varepsilon\hspace{-0.45em}/ \,q\hspace{-0.45em}/\, u_b
\ee
where $m_b$ denotes the b-quark pole mass,
$\varepsilon$ and $q$ are the photon
polarization and momentum, $R=(1+\gamma_5)/2$ is the right handed projection
operator, and $n_f$ is the number
of active flavors of a given mass. The superscript~$(2)$ counts
the powers of $\alpha_s$ and
$M=\,$($0,\, m_b\, \mbox{or}\, m_c$) denotes the mass
of the quark running in the loop inserted into the gluon propagator.
The plots in~FIG.\ref{Zdependence} summarize the outcome of our calculation.
It turned out that the massless approximation overestimates
the massive $b$ result by a large factor, and moreover, has the opposite sign.
On the other hand, less pronounced but non-negligible effects were observed 
for the massive $c$-quark case.
\section{Conclusions}
A complete $\order{\alpha_s^2}$ calculation of the matrix elements
$\langle s \gamma|Q_{1,2}|b \rangle $ is crucial to reduce the overall
uncertainty in the current NNLO estimate of the   
${\mathcal B}({\overline{B}}~\to~ X_s\gamma)$. This calculation is being 
pursued in~\cite{O2BSG}. Taking new results for 
the complete NNLO fermionic contribution into account, an enhancement
of $1.1\%$ for $\mu_b=2.5$ GeV is observed in the current 
estimate of the branching ratio.       
\section{Acknowledgments}
We thank the organizers of the RADCOR 2007 conference 
for putting together such a stimulating meeting.
Useful discussions with Mikolaj Misiak are acknowledged.
This work is supported by the Sofja Kovalevskaja Award 
of the Alexander von Humboldt Foundation
sponsored by the German Federal Ministry of Education and Research.

\end{document}